\documentclass[useAMS,usenatbib,usegraphicx]{mn2e}
\usepackage{amsmath}
\usepackage{url}

\title[Very high energy $\gamma$-ray emission from RBS 0679]{Very high energy $\gamma$-ray emission from RBS 0679}

\author[A.M. Brown et al.]{Anthony M. Brown$^{1}$\thanks{E-mail: anthony.brown@durham.ac.uk}, Paula M. Chadwick$^{1}$ and Hermine Landt$^{1}$\\
$^{1}$Department of Physics, University of Durham, South Road, Durham, DH1 3LE, UK}
\begin{document}

\date{Accepted XXX. Received XXX. In original form 2014 June 30}

\pagerange{\pageref{firstpage}--\pageref{lastpage}} \pubyear{2014}

\maketitle

\label{firstpage}

\begin{abstract}
In this paper we report the \textit{Fermi} Large Area Telescope (LAT) detection of Very High Energy (VHE; $E_{\gamma}>100$ GeV) $\gamma$-ray emission from the BL Lac object RBS 0679. 5.3 years of LAT observations revealed the presence of three VHE photon events within 0\ensuremath{^{\circ}}.1 of RBS 0679, with a subsequent unbinned likelihood analysis finding RBS 0679 to be a source of VHE photons at $6.9$ standard deviations ($\sigma$). An unbinned likelihood analysis of the $0.1-100$ GeV data, binned in 28-day periods, finds both the flux and spectral index to be variable, with a `softer-when-brighter' trend in the global $\gamma$-ray characteristics. On the other hand, the 28-day periods in which the VHE photons were detected have spectral indices that are consistent with the 5.3 year average suggesting that the observed VHE emission is not associated with a spectral hardening event. The discovery of RBS 0679 in the $100-300$ GeV energy range, combined with the non-detection above 390 GeV with the H.E.S.S. telescope array, suggest RBS 0679 to be an intriguing source that requires further follow-up observations with ground-based $\gamma$-ray observatories.
\end{abstract}


\section{INTRODUCTION}
Launched in June 2008, the \textit{Fermi} $\gamma$-ray Space Telescope affords an ideal opportunity to investigate the inner workings of active galactic nuclei (AGN). To date, \textit{Fermi} has spent over 95\% of its lifetime in \textit{all-sky-survey} observing mode, whereby the Large Area Telescope (LAT) onboard \textit{Fermi} points away from the Earth and rocks north and south of its orbital plane on consecutive orbits. This rocking motion of the \textit{Fermi}-LAT detector, coupled with it's large effective area, allows \textit{Fermi} to scan the entire $\gamma$-ray sky every two orbits, or approximately every three hours (\citet{atwood}). The scanning ability of the LAT has allowed us to catch AGN during brief flares of $\gamma$-ray activity (e.g. \citet{dickinson}), with these flares sometimes resulting in the discovery of Very High Energy (VHE; $E_{\gamma}>100$ GeV) emission from the AGN (e.g. \citet{ong} \& \citet{aliu}). 

While its three hour scan period is important for catching brief periods of flare activity from AGN, coupling \textit{Fermi}-LAT's continual scanning of the sky with a long mission lifetime allows us to construct a deep exposure of the extragalactic $\gamma$-ray sky. This deep exposure affords us the ability to perform searches for faint VHE sources which would otherwise would go undetected by the pointed observations of  ground-based imaging atmospheric cherenkov telescope (IACT) arrays. Utilising such an approach, a recent study of a deep $\sim5.3$ year \textit{Fermi}-LAT exposure uncovered VHE emission from the BL Lac object RBS 0970 (\citet{meVHE1}). 

RBS 0679, also known as BZB J0543-5532, is a point like radio source, at a redshift of z $=0.273$ (\citet{redshift}). Detected as a bright X-ray source with the \textit{ROSAT} satellite (\citet{xray}), RBS 0679 has been optically identified as a BL Lac object (\citet{opt}). Discovered as a source of $\gamma$-rays in \textit{Fermi}-LAT observations (\citet{gamma}), RBS 0679 is included in both the 1 and 2 year LAT AGN catalogues (1LAC, \cite{1lac}; 2LAC \cite{acker3}), as well as the recent $E_{\gamma}>10$ GeV LAT catalogue (\citet{hecat}). Recently the H.E.S.S. collaboration has published an upper limit on RBS 0679 of $9\times10^{-13} \text{ cm}^{-2} \text{s}^{-1}$, from 8.1 hours of observations (\citet{hess2}). The energy threshold for this upper limit was 390 GeV. Interestingly, the H.E.S.S. observations found that, even after accounting for absorption of the $\gamma$-ray flux by the extragalactic background light (EBL), this upper limit is not compatible with a simple power-law extrapolation of RBS 0679's 2LAC spectrum to above 390 GeV. This suggests the presence of an intrinsic break in the $\gamma$-ray spectrum, not associated with the process of EBL absorption.

This paper reports the discovery of VHE emission from RBS 0679. Using the same approach that discovered RBS 0970 as a VHE source, 5.3 years of \textit{Fermi}-LAT data also revealed three \textsc{ultraclean} $E_{\gamma}>100$ GeV events clustered within 0\ensuremath{^{\circ}}.1 of RBS 0679. Unlike RBS 0970, the VHE emission detected from RBS 0679 does not coincide with severe spectral hardening, with the spectral index of the $0.1-100$ GeV flux during the detection of the VHE photons being consistent with the 5.3 year average. However, the F$_{\gamma} - \Gamma$ parameter space suggests a softer-when-brighter property for the global $\gamma$-ray characteristics throughout the 5.3 year data set. In \textsection 2 the \textit{Fermi}-LAT observations and analysis routines used in this study are described, along with the results of the $0.1-300$ GeV likelihood analysis. The results of the VHE study of RBS 0679 are shown in \textsection 3. A brief investigation into the global $\gamma$-ray characteristics of RBS 0679 when the VHE emission occurs is presented in \textsection 4, with the conclusions given in \textsection 5. It should be highlighted that this paper reports the discovery of VHE emission from RBS 0679. A follow-up publication with SED modelling and interpretation is currently in preparation and will follow shortly.

\section{\textit{Fermi}-LAT OBSERVATIONS AND DATA ANALYSIS}
The data used in this study comprises all \textit{Fermi}-LAT event and spacecraft data taken during the first 5.3 years of \textit{Fermi}-LAT operation, from 2008 August 4 to 2013 December 12, which equates to a Mission Elapsed Time (MET) interval of 239557417 to 408871812. All \textsc{source} $\gamma$-ray events\footnote{\textsc{source} events have an event class of 2 in the \textsc{pass}7\_\textsc{rep} data, and have a high probability of being a $\gamma$-ray (see \cite{acker2} for details on event classification).}, in the $0.1 < E_{\gamma} < 300$ GeV energy range, within a 13\ensuremath{^{\circ}} radius of interest (RoI) centered on the Second Fermi AGN Catalogue (2LAC; \cite{acker3}) position of RBS 0679, ($\alpha_{J2000}$, $\delta_{J2000}=85$\ensuremath{^{\circ}}.987, $-55$\ensuremath{^{\circ}}.5344), were considered. In accordance with the \textsc{pass}7\_\textsc{rep} criteria, a zenith cut of 100\ensuremath{^{\circ}} was applied to the data to remove any cosmic ray induced $\gamma$-rays from the Earth's atmosphere. The good time intervals were generated by applying a filter expression of ``\textsc{(data\_qual$==$1) \&\& (lat\_config$==$1) \&\& abs(rock\_angle)$<$ 52}'' to the data, where the \textsc{(data\_qual)} and \textsc{(lat\_config)} flags remove sub-optimal data affected by spacecraft events and the \textsc{(abs(rock\_angle))} flag removes data periods where the LAT detector rocking is greater than 52\ensuremath{^{\circ}}\footnote{See \url{http://fermi.gsfc.nasa.gov/ssc/data/analysis/documentation/Cicerone/Cicerone_Data_Exploration/Data_preparation.html} for details on LAT data selection.}. 

Throughout this analysis, version \textsc{v9r32p5} of the \textit{Fermi Science Tools} was used in conjunction with the \textsc{p7rep\_clean\_v15} and \textsc{p7rep\_source\_v15} instrument response functions (IRFs). The IRFs defines how the LAT detector performs for various parameters such as incident photon energy, incident angle and location of photon conversion within the detector. During the analysis, a model file consisting of both point and diffuse sources of $\gamma$-rays was employed. In particular, the model file consisted of the most recent galactic and extragalactic\footnote{For the galactic diffuse model, we have used the LAT collaboration's `gll\_iem\_v05.fit' model. For the extragalactic diffuse, we have used both the `iso\_clean\_v05.txt' and `iso\_source\_v05.txt' extragalactic diffuse models, depending upon whether source or clean photon events were considered. These models are available at \url{http://fermi.gsfc.nasa.gov/ssc/data/access/lat/BackgroundModels.html}.} diffuse models, and all $\gamma$-ray point sources within a 18\ensuremath{^{\circ}} RoI centered on RBS 0679. The positions of these point sources, along with their spectral shapes, were taken from the Second Fermi Source Catalogue (2FGL; \cite{nolan}). Located 13\ensuremath{^{\circ}}.03 from RBS 0679, the extended $\gamma$-ray emission associated with the Large Magellanic Cloud was accounted for with the \textit{Fermi}-LAT collaboration's \textsc{lmc.fits} spatial model, with the normalisation and spectral parameters frozen to their 2FGL values. The normalisation factor of the extragalactic diffuse emission was left free to vary, while the galactic diffuse template was multiplied by a power law in energy, the normalisation of which was left free to vary.

Firstly a binned maximum likelihood analysis was performed on the entire 5.3 year data set in the $0.1-300$ GeV energy range. For RBS 0679 itself, a power law spectral shape of the form $dN/dE = $ A$ \times (E/E_o)^{-\Gamma}$ was assumed, with the normalisation, A, and the spectral index, $\Gamma$, left free to vary. The normalisation and spectral parameters of all point sources within 13\ensuremath{^{\circ}} of RBS 0679 were left free to vary, while the normalisation and spectral parameters for all point sources within an annulus of 13\ensuremath{^{\circ}} to 18\ensuremath{^{\circ}} from RBS 0679 were fixed to those published in the 2FGL catalogue.  

Utilising the above described model, the binned likelihood analysis of the 5.3 year data set, utilising the \textsc{p7rep\_source\_v15} IRF, resulted in the following best-fit power law function for RBS 0679:

\begin{equation}
 \dfrac{dN}{dE}= (1.92 \pm 0.12) \times 10^{-13} (\dfrac{E}{3006.4\text{ MeV}})^{-1.82\pm0.05} \nonumber
\end{equation}

\begin{equation}
 \text{ photons cm}^{-2} \text{s}^{-1} \text{MeV}^{-1}
\end{equation}

which equates to an integrated flux, in the $0.1-300$ GeV energy range, of

\begin{equation}
  F_{0.1<E<300\text{ GeV}} = (1.14 \pm 0.14) \times 10^{-8}  \text{ photons cm}^{-2} \text{s}^{-1}
\end{equation}

taking only statistical errors into account\footnote{Primarily governed by the uncertainty in the effective area, the systematic uncertainty of the integrated flux is energy dependent and is currently estimated as 10\% at 100 MeV, down to 5\% at 560 MeV and back to 10\% for 10 GeV photons (\citet{acker1}).}. For the best-fit power law description, a test statistic\footnote{The test statistic, TS, is defined as twice the difference between the log-likelihood of two different models, $TS=2[\text{log} L - \text{log} L_{0}]$, where $L$ and $L_{0}$ are defined as the likelihood when the source is included or not respectively (\citet{mattox2}).} of $TS=668$ was found, corresponding to a $\sim26\sigma$ detection of RBS 0679 in the $0.1-300$ GeV energy range over the 5.3 year period. 

The long exposure of the analysed observations can result in additional faint sources being present in the data that are not present in the 2FGL. If these sources are present in the data, and not properly accounted for within the model file utilised during the likelihood analysis, they can artifically increase the significance of the $\gamma$-ray flux from RBS 0679 (e.g. \citet{mepicA}, \citet{oscar} \& \citet{meVHE1}). To check if indeed any additional sources were present, a residual count map was constructed in the $0.1-300$ GeV energy range. Shown in Figure 1, the residuals map was obtained by subtracting a `model' map\footnote{The model map was created with the \textsc{gtmodel} \textit{Fermi} tool in conjuction with the best-fit model of the $0.1-300$ GeV likelihood fit of Equation 1.} from the observed counts map, and then dividing by the model map again. The residuals map shows fluctuations in the count residuals at the $\pm 1$\% level. These fluctuations are consistent with random fluctuations and as such, the model used during the analysis is an accurate representation of the observed $\gamma$-rays, with no new $\gamma$-ray point sources being present.

\begin{figure}
 \centering
\includegraphics[width=70mm]{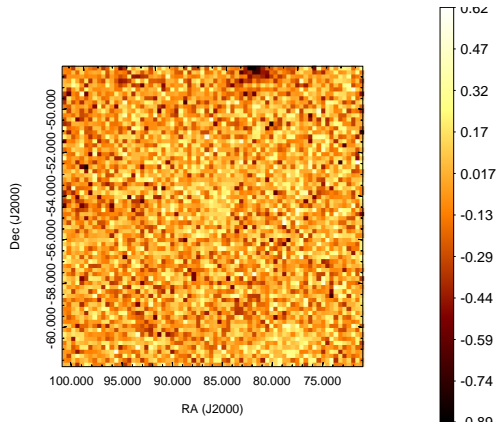}
\caption{A residuals map, ((counts$-$model)$/$model), in the $0.1-300$ GeV energy range, during the entire 5.3 year period, centered on the co-ordinates of RBS 0679. The colour scale is in units of percentage, with fluctuations in the residuals between the model and observed counts map being at the $\pm 1$\% level. These fluctuations are consistent with random fluctuations and as such, show the model used during the analysis to be a good representation of the data. It is worth highlighting that the biggest percentage difference between the model and observed count maps, seen at the top of the residuals map, is positionally coincident with the 2LAC source PMN J0531-4827.}
\label{residuals}
\end{figure}

\section{Very High Energy $\gamma$-ray properties}

\begin{table*}
 \begin{minipage}{140mm}
   \caption{Summary of the three VHE events from RBS 0679 detected by \textit{Fermi}-LAT. It should be noted that all three of these events are also \textsc{ultraclean} class events. The \textsc{gtsrcprob} probabilities refer to the likelihood that the individual VHE photons originated from RBS 0679, as opposed to other sources such as the galactic or extragalactic diffuse emission and have been calculated with the \textit{Fermi} tool \textsc{gtsrcprob}.}
   \begin{center}
     \begin{tabular}{cccccc} \hline \hline
      Energy & MET  & MJD     & $\alpha_{J2000}$ & $\delta_{J2000}$ &  \textsc{gtsrcprob} \\ 
      (GeV)    &   (second) & (day)         & (deg)        & (deg)         & probability \\ \hline
      167    & 253636097.775 & 54845.60373 & 85.972      & -55.502       & 0.999865   \\ 
           &    &        &      &              &             \\
      136    & 293442157.552 &  55306.32201 & 85.994      & -55.561   & 0.999909   \\ 
           &    &        &      &              &             \\
      257   & 355339624.208  & 56022.72787 & 85.959      & -55.528       & 0.99998    \\   \hline \hline
    \end{tabular}
  \end{center}
  \label{photondetails}
\end{minipage}
\end{table*}

A closer inspection of the individual photon events within 0\ensuremath{^{\circ}}.1 of RBS 0679 revealed the presence of three VHE $\gamma$-rays. All three $\gamma$-rays are classed as \textsc{ultraclean} events\footnote{See \cite{acker2} for details on event classification.}, a subclass of \textsc{clean} events that have the highest probability of being photons. Using the combined diffuse and point source model file from \textsection 2, with all normalisation and spectral parameters frozen to the best-fit values of the 5.3 year binned likelihood analysis, the \textsc{gtsrcprob} \textit{Fermi} tool was used to calculate the probability that each of the VHE $\gamma$-ray events originated from RBS 0679, as opposed to other sources such as the galactic or extragalactic diffuse emission. The results of these probability calculations are shown in Table \ref{photondetails}, along with the energy, time detected and ($\alpha_{J2000}$, $\delta_{J2000}$) of each photon event.

Considered in isolation, none of these $E_{\gamma}>100$ GeV events are significant enough to consider RBS 0679 a source of VHE $\gamma$-rays, the most significant event being the 257 GeV photon detected on MJD$=56022.72787$, with a $>4\sigma$ confidence of originating from RBS 0679, assuming Gaussian errors. The other two VHE events have a $>3\sigma$ confidence of originating from RBS 0679. However, the clustering of such energetic photons within a relatively small area can be significant given the small background rates detected by the \textit{Fermi}-LAT above 100 GeV (e.g. \citet{neronov} \& \citet{tanaka}). To determine if RBS 0679 is indeed a source of VHE $\gamma$-ray photons, an unbinned likelihood analysis, with the \textsc{p7rep\_clean\_v15} IRF, was applied to all $E_{\gamma}>100$ GeV \textsc{clean}\footnote{\textsc{clean} events have an event class of 3 in the \textsc{pass}7\_\textsc{rep} data, (see \cite{acker2} for details on event classification).} events within 5\ensuremath{^{\circ}} of RBS 0679 for the entire 5.3 year data set. To accommodate the smaller RoI, the model file used only considered point sources within 6\ensuremath{^{\circ}} of RBS 0679. The normalisation and spectral parameters of point sources within 5\ensuremath{^{\circ}} were left free to vary, while point sources within the 5\ensuremath{^{\circ}} to 6\ensuremath{^{\circ}} annulus had their normalisation and spectral parameters frozen to their 2FGL values. The diffuse components were treated the same as in \textsection 2, with the exception that the extra-galactic diffuse was modelled with the iso\_clean\_v05.txt description. The resultant best-fit power law function for RBS 0679 in the $100-300$ GeV energy range, was found to be:

\begin{equation}
 \dfrac{dN}{dE}= (0.06 \pm 0.3) \times 10^{-12} (\dfrac{E}{3006.4\text{ MeV}})^{-0.44\pm1.53} \nonumber
\end{equation}

\begin{equation}
 \text{ photons cm}^{-2} \text{s}^{-1} \text{MeV}^{-1}
\end{equation}

which equates to an integrated flux of

\begin{equation}
  F_{E>100\text{ GeV}} = (1.98 \pm 1.15) \times 10^{-11}  \text{ photons cm}^{-2} \text{s}^{-1}
\end{equation}

again taking only statistical errors into account. It is worth highlighting that due to the limited number of photons with $E_{\gamma}>100$ GeV, the statistical errors of the likelihood fit are large. The TS value of the best-fit power law was $TS=47.5$, equating to a significance of $\sim6.9\sigma$. As such, this analysis represents the discovery of RBS 0679 as a source of VHE $\gamma$-rays. 

To locate the origin of the VHE $\gamma$-ray emission, another \textit{Fermi} tool, \textsc{gtfindsrc}, was applied to all $E_{\gamma}>100$ GeV events within 5\ensuremath{^{\circ}} of RBS 0679. Using the same combined diffuse and point source model that was applied during the unbinned likelihood fit to the VHE data, the observed VHE $\gamma$-ray emission was found to originate from ($\alpha_{J2000}$, $\delta_{J2000}=$ 85\ensuremath{^{\circ}}.9734, -55\ensuremath{^{\circ}}.535), with a 95\% error radius of 0\ensuremath{^{\circ}}.025. The VHE emission is therefore found to be spatially co-incident with the 2LAC position of RBS 0679. 

\section{Discussion}

With our discovery of VHE emission positionally coincident with RBS 0679, we firstly turn our attention to the apparent discrepancy between our result and the H.E.S.S. upper limit. Extrapolating our best-fit power law model from \textsection 3, and correcting for EBL extinction with the \citet{fran} EBL model, we find that, within errors, there is no discrepancy between the two analyses. We do however note that there is a large uncertainty in the spectral index of our best-fit power law model. This large uncertainty, understandable given the small number of photons that the LAT has detected in the VHE regime, limits our ability to investigate the presence and nature of a cut-off in RBS 0679's VHE spectrum. 

Investigating the presence and nature of a cut-off in the $\gamma$-ray spectrum of RBS 0679 requires detailed SED modelling studies, in conjunction with observations of RBS 0679 with the recently commissioned HESS-II telescope. Detailed SED modelling studies of RBS 0679's broad-band emission are currently on-going, the results of which will be presented in a future publication. Along with allowing us to better understand the inner workings of RBS 0679, these SED studies will allow us to better determine when to trigger HESS-II observations. With an energy threshold of $\sim30$ GeV, HESS-II will allow us to observe both the high-energy component of the LAT spectrum and the low energy component of the H.E.S.S. spectrum with greatly improved statistics. These improved photon statistics will allow us to determine conclusively if a cut-off in the spectrum is present. 

\begin{figure*}
 \centering
 \begin{minipage}{170mm}
\includegraphics[width=170mm]{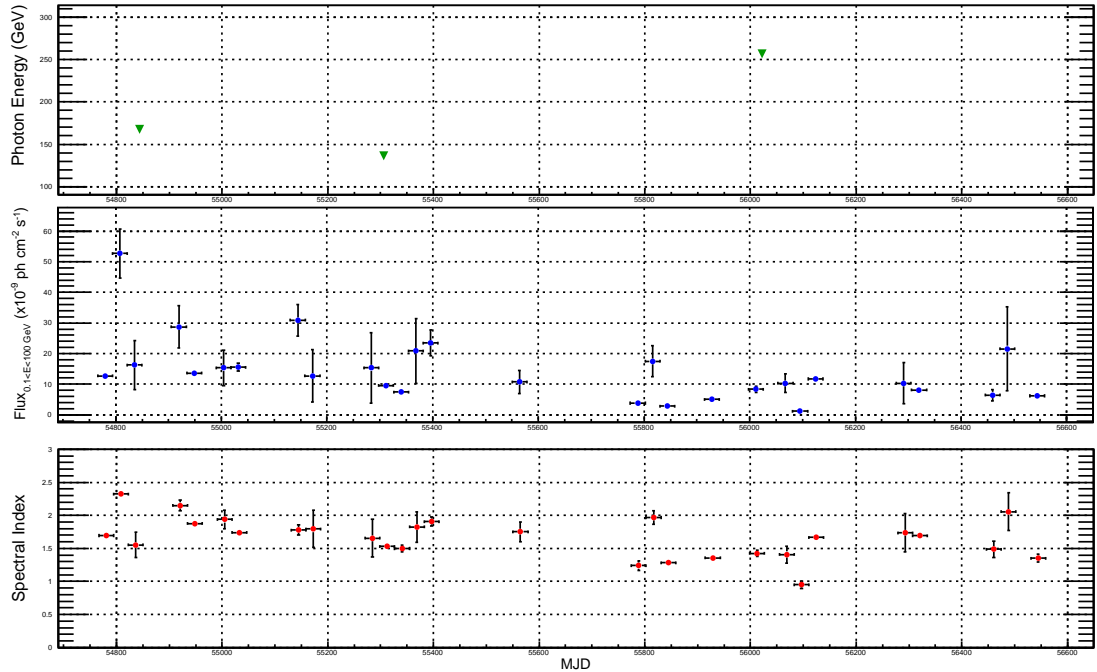}
\caption{\textit{Top panel}: Arrival times in MJD of all $E_{\gamma}>100$ GeV \textsc{clean} events within 0\ensuremath{^{\circ}}.1 of RBS 0679. \textit{Middle panel}: $0.1-100$ GeV flux light curve, binned in 28-day periods. Only bins where the TS is greater than 10 are shown. \textit{Bottom panel}: Spectral index for each 28 day period where the $TS \geq 10$.}
\label{lc}
\end{minipage}
\end{figure*}
To investigate the global $\gamma$-ray properties of RBS 0679 during the emission of the VHE $\gamma$-rays, the temporal characteristics of the $\gamma$-ray flux and spectral index were investigated. The entire 5.3 year data set was binned into 28-day periods, with an unbinned likelihood analysis applied to each bin separately. The 28-day bin size was chosen simply as a compromise between having decent statistics in each bin, while being able to resolve structure within the lightcurve. All \textsc{source} events in the 0.1 $< E_{\gamma} <$ 100 GeV energy range were considered.  The upper energy limit of 100 GeV was chosen so as to remove any possible bias to harder spectral indices caused by the presence of the VHE events. The good time intervals were selected as outlined in \textsection 2. The model file utilised in each \textsc{gtlike} fit was the same as that used in \textsection 2; that is, the normalisation and spectral parameters of all point sources within 13\ensuremath{^{\circ}} of RBS 0679 were left free to vary, while the normalisation and spectral parameters for all point sources within an annulus of 13\ensuremath{^{\circ}} to 18\ensuremath{^{\circ}} from RBS 0679 were fixed to those published in the 2FGL . Only time intervals where the corresponding TS value was greater than 10 were considered, which equates to a significance of $\approx 3 \sigma$. The resultant flux and spectral index `lightcurve' can be seen in Figure 2, along with the arrival times of all $E_{\gamma}>100$ GeV photons within 0\ensuremath{^{\circ}}.1 of RBS 0679. 

\citet{kat} and \citet{mengc} found that, for NGC 1275, it is the $E_{\gamma}\geq1$ GeV $\gamma$-ray flux and spectral shape that are important when triggering ground-based VHE $\gamma$-ray observations, with a higher $E_{\gamma}\geq1$ GeV flux, or harder $\gamma$-ray spectrum, more likely to be associated with the emission of VHE $\gamma$-ray photons. Likewise, the majority of VHE photons detected from RBS 0970 coincided with severe spectral hardening (\citet{meVHE1}). However, this `harder-when-brighter' phenomenon for the VHE emission is not universally applicable to all VHE emitting AGN, with \citet{mepks} finding no such trend for the FSRQ PKS 1510-089. The variable nature of RBS 0679's flux and spectral index, shown in Figure 2, is clear to see. Indeed, a constant flux fit to the flux distribution of Figure 2 has a $\chi_{red}^{2}$ of 289.9, while a constant index fit to the spectral indices distribution of Figure 2 has a $\chi_{red}^{2}$ of 3337.7. Interestingly, in the F$_{\gamma}-\Gamma$ parameter space, shown in Figure 3, this flux and spectral variability shows a clear `softer-when-brighter' trend. Such a trend has previously been seen in some AGN in \textit{EGRET} data (eg. \citet{nan}). More recently the H.E.S.S. collaboration has seen both `softer-when-brighter' and `harder-when-brighter' trends from the prominent BL Lac object PKS 2155-304 in the VHE regime, depending upon the flux level (\citet{hess}). 

In the traditional synchrotron self-compton\footnote{The $\gamma$-ray flux from BL Lac object subclass of AGN is often attributed to the synchrotron self-compton (SSC) model, whereby the observed $\gamma$-ray flux is produced through the inverse comptonisation of synchrotron photons by a population of relativistic electrons (eg. \citet{kraw}; \citet{mephd}; \citet{abram}).} model, a `softer-when-brighter' trend can be reproduced by an increase in the magnetic field strength, a softening of the primary electron population or an increase in the size of the emission region and the associated adiabatic cooling. Outside of the SSC interpretation, a `softer-when-brighter' characteristic suggests a rapid cooling of the highest energy particles as RBS 0679 brightens. To determine which of these effects, if any, are responsible for the observed `softer-when-brighter' characteristic of RBS 0679 requires indepth multi-wavelength studies which we defer to a later paper. 

At first glance, Figure 3 suggests that RBS 0679 is not necessarily VHE bright during the harder spectral states. Considering the VHE photons on an individual basis, this does indeed appear to be true, with all three VHE photons being detected during months where the spectral index was consistent, within errors, with the 5.3 year average (see Table 2). Nonetheless, it is worth noting that phenomenological studies by \textit{Fermi} have found the majority of TeV bright AGN have a $\Gamma < 2$ in the $0.1-100$ GeV energy range (\citet{abdo}). As such, while the emission of VHE photons from RBS 0679 does not appear to be associated with any spectral hardening phenomenon, the monthly spectral indices of RBS 0679 during the detection of the VHE $\gamma$-rays do follow the trend that VHE bright AGN have a $\Gamma < 2$ in the \textit{Fermi}-LAT energy range.

With the detection of three \textsc{ultraclean} events within 0.1\ensuremath{^{\circ}}, the VHE detection of RBS 0679 is a robust result. As such, with a redshift of z=0.273, RBS 0679 is a relatively distant, spectrally hard, VHE emitting BL Lac object. AGN with these characteristics are ideal for studying the intensity of the EBL (\citet{coppi}) and the strength of the intergalactic magnetic field (IGMF; \citet{neronov3}). As such, besides better understanding the VHE properties of RBS 0679, ground-based observations with IACTs will also allow us to use RBS 0679's $\gamma$-ray spectrum to study the EBL and determine the suitability of RBS 0679 for studying the IGMF.

\begin{figure}
 \centering
\includegraphics[width=70mm]{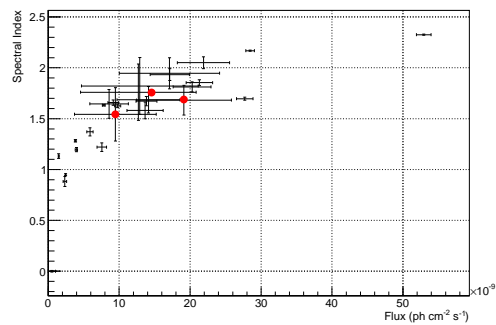}
\caption{The F$_{\gamma}-\Gamma$ parameter space of the individual 28 day intervals shown in Figure 2. The F$_{\gamma}-\Gamma$ parameter space shows a clear `softer-when-brighter' tendency for the global $\gamma$-ray characteristics of RBS 0679. The red points represent the F$_{\gamma}-\Gamma$ values for the 28 day period in which a VHE photon was detected.}
\label{specvsflux}
\end{figure}

\begin{table}
   \caption{Summary of the global $\gamma$-ray properties of RBS 0679, in the $0.1-100$ GeV energy range, during the 28 day period in which the VHE photons were detected, along with 5.3 year average values. The photon energy is in units of GeV, Flux and $\Delta$Flux are in units of $\times10^{-9}$ ph cm$^{-2}$ s$^{-1}$ and are the $0.1-100$ GeV flux and statistical error of the flux. $\Gamma$ and $\Delta \Gamma$ are the spectral index and statistical error of the spectral index.}
   \begin{center}
     \begin{tabular}{ccccc} \hline \hline
      Photon Energy & Flux & $\Delta$Flux & $\Gamma$ & $\Delta \Gamma$ \\  \hline
      167    & 19.13 & 6.69 & 1.68  & 0.15              \\ 
      136    & 14.56 & 0.39  & 1.75  & 0.01              \\ 
      257    & 9.47 & 5.76 & 1.54  & 0.26              \\   \hline 
      5.3 year average & 11.27 & 1.56 & 1.81 & 0.06 \\ \hline \hline
    \end{tabular}
  \end{center}
  \label{indexdetails}
\end{table}

\section{Conclusions}

With 5.3 years of \textit{Fermi}-LAT data, RBS 0679 has been found to be a source of VHE $\gamma$-ray photons. With three \textsc{ultraclean} $E_{\gamma}>100$ GeV photon events within 0.1\ensuremath{^{\circ}} of RBS 0679, an unbinned likelihood analysis revealed the significance of this discovery to be at the $6.9\sigma$ confidence level. The 5.3 year integrated $E_{\gamma}>100$ GeV flux was found to be $(1.98 \pm 1.15) \times 10^{-11}  \text{ photons cm}^{-2} \text{s}^{-1}$. 

An investigation of the global $0.1-100$ GeV $\gamma$-ray characteristics during the 5.3 years revealed a `softer-when-brighter' trend. Interestingly, the 28-day periods during which the VHE photons were detected had a spectral index that was consistent with the 5.3 year average, suggesting that the observed VHE emission is not associated with a spectral hardening event.

The detection of three \textsc{ultraclean} events within 0.1\ensuremath{^{\circ}} of RBS 0679, coupled with the results of the $100-300$ GeV unbinned likelihood analysis, suggest the VHE detection of RBS 0679 is a robust result. While the extrapolation of RBS 0679's spectrum, as reported in previous \textit{Fermi}-LAT catalogues, is inconsistent with the recently announced H.E.S.S. upper limit, the extrapolation of the best-fit power law for the VHE discovered here is consistent with the H.E.S.S. upper limit. However, given the large uncertainty on the spectral index of this power-law fit does not significantly rule out the presence of a cut-off in the $\gamma$-ray spectrum. To investigate the presence of a cut-off in detail requires follow-up observations with IACT arrays such as HESS-II or the future CTA (\citet{cta}), in conjunction with indepth SED modelling. This indepth SED modelling is currently ongoing, the results of which we defer to a later publication.

\section*{Acknowledgments}

We thank the referee for their comments and suggestions that has improved the quality of this paper. This work was undertaken with the financial support of Durham University and has made use of public \textit{Fermi} data obtained from the High Energy Astrophysics Science Archive Research Center (HEASARC), provided by NASA’s Goddard Space Flight Center. This work has also made use of the NASA/IPAC Extragalactic Database (NED), which is operated by the Jet Propulsion Laboratory, Caltech, under contract with the National Aeronautics and Space Administration. We thank the \textit{Fermi}-LAT collaboration for the quality of the data and analysis tools that were used in this study.

\end{document}